\documentclass[portrait]{seminar}
\usepackage{amsmath,amssymb}
\numberwithin{equation}{section}
\newtheorem{definition}{DEFINITION}[section]
\newtheorem{theorem}{Theorem}[section]

\newtheorem{remark}{Remark}[section]
\newtheorem{axiom}{AXIOM}[section]
\newtheorem{constraint}{CONSTRAINT}[section]
\newtheorem{diagram}{DIAGRAM}[section]
\newenvironment{hypothesis}{HP: \begin{center}} {\end{center}}
\newenvironment{thesis}{TH: \begin{center}} {\end{center}}
\newenvironment{proof}{\begin{center}PROOF: \end{center}} {
$ \blacksquare $}

\newtheorem{example}{Example}[section]
\title{THE DEFINITION OF A RANDOM SEQUENCE OF QUBITS: FROM NONCOMMUTATIVE ALGORITHMIC PROBABILITY
THEORY TO QUANTUM ALGORITHMIC INFORMATION THEORY AND BACK}
\author{Gavriel Segre - University of Pavia, Europe}
\begin{document}
\begin{slide*}
\maketitle
\end{slide*}
\begin{slide*}
\textbf{ACKNOWLEDGMENTS:}

First of all I want  to thank :
\begin{itemize}
  \item Asterix
  \item F. Benatti
  \item C. Calude
  \item G. Jona-Lasinio
  \item Obelix
  \item P. Odifreddi
  \item M. Rasetti
  \item A. Rimini
  \item K. Svozil
  \item M. Van Lambalgen
\end{itemize}
for useful discussions and  suggestions.

They all have no responsibility for any mistake contained in these
pages.
\end{slide*}
\begin{slide*}
\tableofcontents
\end{slide*}
\begin{slide*}
\section{Introduction}
Equivalent approaches to the definition of a random sequence over
a \textbf{(commutative) finite alphabet $ \Sigma $  }:
\begin{itemize}
  \item \textit{Chaitin's definition \cite{Chaitin-66}, \cite{Chaitin-69}, \cite{Chaitin-87}, \cite{Calude-94}, \cite{Li-Vitanyi-97}:}

    \textbf{algorithmic incompressibility} in the framework of
  \textbf{(Commutative) Algorithmic Information Theory}

  \item \textit{definition by Martin-L\"{o}f tests \cite{Martin-Lof-66a}, \cite{Martin-Lof-66b}, \cite{Chaitin-87}, \cite{Calude-94}, \cite{Li-Vitanyi-97}:}

   \textbf{passage of all the algorithmically implementable (commutative) statistical tests}
\end{itemize}
\end{slide*}
\begin{slide*}
\begin{itemize}
  \item \textit{Martin-L\"{o}f's algorithmic measure-theoretic definition \cite{Martin-Lof-66a}, \cite{Martin-Lof-66b}, \cite{Chaitin-87}, \cite{Calude-94}, \cite{Li-Vitanyi-97}:}

   \textbf{not belongness to any set of null algorithmic (commutative)
  unbiased probability}

  \item \textit{Solovay's algorithmic measure-theoretic definition \cite{Solovay-77}, \cite{Calude-94}, \cite{Li-Vitanyi-97}}

  \item \textit{some (still lacking!) restriction of Von-Mises-Church's definition \cite{von-Mises-81}, \cite{Church-40}, \cite{Longpre-92}, \cite{Li-Vitanyi-97}:}

   \textbf{stability of the relative-frequencies of the various (commutative) letters} under the extraction of a subsequence by a
   \textbf{properly subset of the (commutative) algorithmic place selection rules}
\end{itemize}
\end{slide*}
\begin{slide*}
Common feauture of all these definitions:

 \emph{THEY CONTAIN THE
TERM \textbf{ALGORITHMIC} AND , THUS, DEPEND ON
\textbf{COMPUTABILITY THEORY}}

\emph{This suggest that the same should happen also for the
definition of a random sequence on a \textbf{noncommutative
finite alphabet $ \Sigma_{NC} $}}
\end{slide*}
\begin{slide*}
Conceptual meaning of the inelusibility of Computability Theory:

\emph{\textbf{COMMUTATIVE MEASURE THEORY} can't resolve by itself
the definition of a random sequence on a commutative alphabet
suggesting the requirement of an alternative \textbf{ALGORITHMIC
FOUNDATION OF COMMUTATIVE PROBABILITY THEORY} deeply pursued by
the same father of the measure-theoretic foundation A.N.
Kolmogorov \cite{Shiryayev-94}}

\emph{This suggest that the same should be true as to
\textbf{NONCOMMUTATIVE PROBABILITY} leading to the idea of
pursuing an \textbf{ALGORITHMIC FOUNDATION OF NONCOMMUTATIVE
PROBABILITY THEORY}}
\end{slide*}
\begin{slide*}
The individuation of the \textbf{correct noncommutative
generalization of Martin-L$\ddot{o}$f definition} should be
equivalent to the characterization of a random sequence on a
noncommutative alphabet as \textbf{algorithmic incomprimible} in
the framework of  \textbf{Quantum Algorithmic Information Theory}
\cite{Svozil-96}, \cite{Manin-99},\cite{Vitanyi-99},
\cite{Berthiaume-van-Dam-Laplante-00} giving some light on the
nature of such a theory.
\end{slide*}
\begin{slide*}
\section{Strings and sequences over commutative and noncommutative alphabets}

Given the commutative alphabet of one cbit $ \Sigma\equiv \{ 0,1
\} $ :
\begin{definition}
\end{definition}
SET OF THE STRINGS ON $ \Sigma $ :
\begin{equation}
\Sigma^{\star} \; \equiv \; \cup_{ k \in {\mathbb{N}}} \Sigma^{k}
\end{equation}
\begin{definition}
\end{definition}
SET OF THE SEQUENCES ON $ \Sigma $ :
\begin{equation}
\Sigma^{\infty} \; \equiv \; \{ \bar{x} : {\mathbb{N}}_{+} \,
\rightarrow \, \Sigma \}
\end{equation}
\end{slide*}
\begin{slide*}
\begin{theorem} \label{th:cardinalities of strings and sequences}
\end{theorem}
(ON THE CARDINALITIES OF STRINGS AND SEQUENCES)
\begin{align}
  cardinality(\Sigma^{\star}) \; & = \; \aleph_{0}   \\
  cardinality(\Sigma^{\infty}) \; & = \;  \aleph_{1}
\end{align}
\begin{remark}
\end{remark}
ON THE ASSUMPTION OF NOT INTERMEDIATE DEGREES OF INFINITY BETWEEN
$\Sigma^{\star}$ AND  and $\Sigma^{\infty}$

I will assume from now on the following:
\begin{axiom}
\end{axiom}
CONTINUUM HYPOTHESIS:
\begin{equation}
  2^{\aleph_{0}} \; = \; \aleph_{1}
\end{equation}
that is well known to be \textbf{consistent} but
\textbf{independent} from the formal system of Zermelo - Fraenkel
endowed with the Axiom of Choice (ZFC) giving foundation to
Mathematics \cite{Odifreddi-89}
\end{slide*}
\begin{slide*}
\begin{definition}
\end{definition}
DIADIC EXPANSION:

\begin{equation}
\begin{split}
   de & : \Sigma^{\infty}  \rightarrow [ 0 , 1 ]  \\
   de & ( x_{1} , x_{2}, \ldots ) = \sum_{n=1}^{\infty} \frac{x_{n}}{2^{n}}
\end{split}
\end{equation}

\begin{remark}
\end{remark}
NOT BIJECTIVITY OF THE DIADIC EXPANSION:

de is injective but not surjective since each point of the closed
unitary interval has two counter images: one \emph{terminating}
and one \emph{nonterminating}; e.g.:
\begin{equation}
  de^{-1} ( \frac{1}{2} ) \; = \; \{ 100000 \cdots \, ,
  \, 01111 \cdots \}
\end{equation}
\end{slide*}
\begin{slide*}
\begin{definition}
\end{definition}
CYLINDER SET W.R.T. $ \vec{x} \, =  ( x_{1} , \ldots , x_{n} ) \,
\in \, \Sigma^{\star} $:
\begin{multline} \label{eq:cylinder set}
\Gamma_{\vec{x}} \; \equiv \; \{ \bar{y} = ( y_{1} , y_{2} ,
\ldots ) \in \Sigma^{\infty} : \\
 y_{1} = x_{1} , \ldots , y_{n} = x_{n} \}
\end{multline}
\begin{definition}
\end{definition}
CYLINDER - $ \sigma $ - ALGEBRA ON $ \Sigma^{\infty} $:
\begin{equation}
  {\mathcal{F}}_{cylinder} \; \equiv \;  \sigma \text{-algebra generated
  by } \{ \Gamma_{\vec{x}} \, : \, \vec{x} \in \Sigma^{\star} \}
\end{equation}
\begin{definition}
\end{definition}
LEBESGUE UNBIASED PROBABILITY MEASURE ON $ \Sigma^{\infty} $ :
\begin{equation}
  P_{unbiased} ( A ) \;  \equiv \; \mu_{Lebesgue} ( de ( A ) ) \;
  \; A \, \in \,  {\mathcal{F}}_{Borel}
\end{equation}
\end{slide*}
\begin{slide*}
\begin{remark}
\end{remark}
THE UNBIASED PROBABILITY SPACE OF ALL THE SEQUENCES OF CBITS AS
DIRECT PRODUCT OF UNBIASED PROBABILITY SPACES EACH FOR EVERY
SINGLE CBIT:

The unbiased probability space $ ( \Sigma^{\infty} , P_{unbiased}
)  $  of all the sequences of cbits may be expressed as:
\begin{equation}
\begin{split}
   ( \Sigma^{\infty} & , P_{unbiased})  \; = \; \times_{n \in {\mathbb{Z}}} ( \Sigma , C_{\frac{1}{2} , \frac{1}{2}} ) \\
  C_{\frac{1}{2} , \frac{1}{2}}  &  (x) \; \equiv \;  \frac{1}{2}
  \; \; x \in \Sigma
\end{split}
\end{equation}
\end{slide*}
\begin{slide*}
\begin{remark}
\end{remark}
THE UNBIASED PROBABILITY SPACE OF ALL THE SEQUENCES OF CBITS AS A
DEGENERATE NONCOMMUTATIVE PROBABILITY SPACE:

By the \textbf{Gelfand isomorphism} the classical probability
space $ ( \Sigma^{\infty} , P_{unbiased} )  $ may be equivalentely
seen as the degenerate \textbf{noncommutative probability space} (
or \textbf{quantum probability space} or
\textbf{$W^{\star}$-algebraic probability space}, or $ \cdots $
\cite{Parthasarathy-92}, \cite{Cuculescu-Oprea-94},
\cite{Meyer-95}, \cite{Ohya-Petz-93},
\cite{Ingarden-Kossakowski-Ohya-97}, \cite{Hiai-Petz-00}) $ (
L^{\infty} ( \Sigma^{\infty} , P_{unbiased} ) , \tau_{unbiased} )
$ where $ \tau_{unbiased} $ is the tracial state on the Von
Neumann algebra \cite{Sunder-87} $  L^{\infty} ( \Sigma^{\infty} ,
P_{unbiased} ) $ defined as:
\begin{equation}
 \tau_{unbiased} ( f ) \; \equiv \; \int_{\Sigma^{\infty}} f(x) \,
 d P_{unbiased} \;
\end{equation}
\end{slide*}
\begin{slide*}
\begin{remark}
\end{remark}
THE KEY METAPHORE OF NONCOMMUTATIVE PROBABILITY THEORY AND THE
NONCOMMUTATIVE ALPHABET OF ONE QUBIT

The key metaphore of Noncommutative Probability Theory consists
in imaging an illusionary noncommutative corrispective of the
Gelfand-Theorem  and looking to a noncommutative probability
space $ ( A , \omega ) $ as  a sort of $ ( L^{\infty} (
SPACE_{NC} , P_{NC} ) \, , \,  \int_{SPACE_{NC}} dP_{NC} ) $.

So the \textbf{one-qubit $W^{\star}-algebra \; M_{2}(
{\mathbb{C}}) $} endowed with some state may be identified as the
set of the properly-smooth functions over the
\textbf{NONCOMMUTATIVE ALPHABET OF ONE CBIT} : $ \Sigma_{NC} \;
\equiv \{ 0,1 \}_{NC} $
\end{slide*}
\begin{slide*}
\begin{definition}
\end{definition}
UNBIASED NONCOMMUTATIVE PROBABILITY SPACE ON THE ONE QUBIT
ALPHABET $ \Sigma_{NC} $ :
\begin{equation}
\begin{split}
  ( M_{2}
({\mathbb{C}}) & \, , \, \tau_{2} ) \\
  \tau_{2} & (
  \begin{pmatrix}
    a_{11} & a_{12} \\
    a_{21} & a_{22}
  \end{pmatrix}
   ) \; \equiv \; \frac{1}{2} (  a_{11} +  a_{22} )
\end{split}
\end{equation}
\begin{definition} \label{def:sequences on the noncommutative alphabet}
\end{definition}
SET OF THE SEQUENCES ON $ \Sigma_{NC} $ :

\begin{equation}
  L^{\infty} ( \Sigma_{NC}^{\infty} ) \; \equiv s-closure ( \; \otimes_{n \in {\mathbb{N}}}  M_{2}
({\mathbb{C}}) \; )
\end{equation}
\end{slide*}
\begin{slide*}
$ L^{\infty} ( \Sigma_{NC}^{\infty} ) $ is a
\textbf{$II_{1}$-factor} and thus has a canonical (i.e. finite,
normal and faithful ) trace, namely:
\begin{equation}
  \tau_{unbiased} \; \equiv \; \otimes_{n \in {\mathbb{N}}} \tau_{2}
\end{equation}
\begin{definition}
\end{definition}
UNBIASED NONCOMMUTATIVE PROBABILITY SPACE OF ALL THE SEQUENCES OF
QUBITS: $ ( L^{\infty} ( \Sigma_{NC}^{\infty} ) \; , \;
\tau_{unbiased} ) $
\end{slide*}
\begin{slide*}
\section{The randomness of repeated classical and quantum coin
tossings}\label{sec:classical and quantum coins}
 The correct Martin L\"{o}f - Solovay - Chaitin definition of a random
sequence on $\Sigma$ \cite{Martin-Lof-66a}, \cite{Martin-Lof-66b},
\cite{Solovay-77}, \cite{Chaitin-87}, \cite{Calude-94},
\cite{Li-Vitanyi-97}
 satisfies the following intuitive condition:
 \begin{constraint} \label{con:independent sequence of tosses of a classical coin}
 \end{constraint}
\textsf{ON THE NOTION OF A RANDOM SEQUENCE ON THE COMMUTATIVE
ALPHABET $ \Sigma $ :}

\emph{Making infinite \textbf{independent} trials of the
experiment consisting on tossing a \textbf{classical coin} we
must obtain a random sequence with probability one}
\end{slide*}
\begin{slide*}
So a reasonable strategy to identify the correct definition of a
random sequence of qubits would consist in:
\begin{itemize}
  \item formulating an analogous constraint in terms of an
  infinite sequence of experiments consisting in tossing a quantum
  coin
  \item identifying the information that such a constraint gives on the correct way of making
  a noncommutative generalization of  Martin-L\"{o}f's
  algorithmic-measure-theoretic definition
\end{itemize}
\end{slide*}
\begin{slide*}
The commutative random variables ${\mathbf{ c_{t_{1}} }}$ and
${\mathbf{ c_{t_{2}} }}$ on the commutative probability space $ (
L^{\infty} ( \Sigma^{\infty} , P_{unbiased} ) , \tau_{unbiased} )
$ representing the results of the classical-coin tossing at times,
respectively, ${\mathbf{t_{1}} }$  and ${\mathbf{t_{2}} }$ are
assumed to be \textbf{independent}:
\begin{multline}
  \tau_{unbiased}( c_{t_{1}}^{n} c_{t_{2}}^{m} ) \; = \\
   \tau_{unbiased}(
  c_{t_{1}}^{n} ) \, \tau_{unbiased}(
  c_{t_{2}}^{m} ) \; \; \forall n,m \, \in \, {\mathbb{N}}
\end{multline}
\end{slide*}
\begin{slide*}
\textbf{Such a condition, anyway, requires that $ c_{t_{1}}$ and $
c_{t_{2}} $  are commuting among themselves }:
\begin{equation}
  [ \, c_{t_{1}} \, , \, c_{t_{2}} \, ] \; = \; 0
\end{equation}
But such a condition can't, clearly, be true for the
noncommutative random variables ${\mathbf{ \tilde{c}_{t_{1}} }}$
and ${\mathbf{ \tilde{c}_{t_{2}} }}$  on the noncommutative
probability space $ ( L^{\infty} ( \Sigma_{NC}^{\infty} ) ,
\tau_{unbiased} ) $ representing the results of  quantum-coin
tossing at times, respectively, ${\mathbf{t_{1}} }$  and
${\mathbf{t_{2}} }$ having any  \textbf{ noncommutative
correlation} among themselves.
\end{slide*}
\begin{slide*}
The natural corrispective of the notion of \textbf{independence}
for two generic  noncommutative random variables \textbf{x} and
\textbf{y} over a noncommutative probability space $ ( A , \omega
) $  is Dan Virgil Voiculescu's notion of \textbf{freeness}
\cite{Hiai-Petz-00} stating that there doesn't exist any
particular \textbf{relation} linking \textbf{x} and \textbf{y}
besides the fact of belonging to the same $ W^{\star} $-algebra
exactly as happens for two generators of a \textbf{free group}.
\begin{remark}
\end{remark}
FREENESS IMPLIES NOT INDEPENDENCE

Since among the excluded particular relations among x and y there
is also the one stating the compatibility of such random
variables, if \textbf{x} and \textbf{y} are \textbf{free} they
can't be \textbf{independent}
\end{slide*}
\begin{slide*}
\begin{definition}
\end{definition}
THE NONCOMMUTATIVE RANDOM VARIABLES \textbf{x} AND \textbf{y} ON
THE NONCOMMUTATIVE PROBABILITY SPACE $ ( A , \omega ) $ ARE FREE:
\begin{multline}
  \forall n \in {\mathbb{N}}  \; , \; \forall i_{1} , \cdots ,
  i_{n} \, \in \, \{ 1 , 2  \} \; : \\
   i(k) \neq i(k+1) ( 1 \leq k \leq n-1 ) \;   \\
  \omega ( a_{1} \cdots a_{n} )   \; = \; 0 \; \; whenever \, a_{k}
  \in A_{i(k)} \; , \\
   \omega (  a_{k} )  \; = \; 0 \, , \, 1 \leq k
  \leq n \\
  A_{1} \; \equiv \;  generated ( x ) \\
  A_{2} \; \equiv \;  generated ( y )
\end{multline}
\end{slide*}
\begin{slide*}
Returning now to the noncommutative random variables ${\mathbf{
\tilde{c}_{t_{1}} }}$ and ${\mathbf{ \tilde{c}_{t_{2}} }}$ on the
noncommutative probability space $ ( L^{\infty} (
\Sigma_{NC}^{\infty} ) , \tau_{unbiased} ) $ representing the
results of the quantum-coin tossing at times, respectively,
${\mathbf{t_{1}} }$  and ${\mathbf{t_{2}} }$ it appears natural to
assume that they are  \textbf{free}.

\begin{remark}
\end{remark}
The notion of \textbf{freeness} is an equivalence relation on the
noncommutative probability space $ ( A , \omega ) $ and thus
extends immediately to an arbitrary number of noncommutative
random variables.
\end{slide*}
\begin{slide*}
It appears then natural to require that the notion of
\textbf{noncommutative algorithmic randomness} we are looking for
obeys the following:
\begin{constraint} \label{con:free sequence of tosses of a quantum coin}
\end{constraint}
\textsf{ON THE NOTION OF A RANDOM SEQUENCE ON THE NONCOMMUTATIVE
ALPHABET $ \Sigma_{NC} $ :}

\emph{Making infinite \textbf{free} trials of the experiment
consisting on tossing a \textbf{quantum coin} we must obtain a
random sequence with noncommutative probability one}
\end{slide*}
\begin{slide*}
\section{Martin-L\"{o}f random sequences over a commutative alphabet}
\begin{definition}
\end{definition}
$n^{th}$ PREFIX OF THE SEQUENCE  $ \bar{x} \in { \Sigma^{\infty}}
  $ :
\begin{equation}
  \vec{x}(n)  \in \Sigma^{n} :
   \exists \, \bar{y}  \in  \Sigma^{\infty} \; : \; \bar{x} \; = \; \vec{x}(n) \, \cdot \, \bar{y}
\end{equation}
\begin{definition}
\end{definition}
SEQUENCES BEGINNING WITH $ S \; \subset \; \Sigma^{\star} $:
\begin{equation}
  S \Sigma^{\infty} \; \equiv \; \{ \bar{x} \in \Sigma^{\infty} \,
  : \, \vec{x}(n) \in S \, , \, n \in {\mathbb{N}}_{+} \, \}
\end{equation}
Endowed $ \Sigma^{\infty} $ with the \textbf{product topology}
 induced by the \textbf{discrete topology} of $ \Sigma $:
\begin{definition}
\end{definition}
$ S \; \subset \; \Sigma^{\infty} $  IS A NULL SET:
\begin{multline}
  \forall \epsilon > 0 , \exists G_{\epsilon} \subset
  \Sigma^{\infty} \; open \; : \\
 ( S \subset G_{\epsilon}) \; and \;  (  P_{unbiased} (  G_{\epsilon} ) \, <
  \, \epsilon )
\end{multline}
\end{slide*}
\begin{slide*}
\begin{definition}
\end{definition}
UNARY PREDICATES ON $ \Sigma^{\infty} $ :
\begin{equation}
  {\mathcal{P}} ( \Sigma^{\infty} ) \; \equiv \; \{ p_{\bar{x}} \, : \, \text{ predicate about } \bar{x} \in \Sigma^{\infty}  \}
\end{equation}
\begin{definition}
\end{definition}
TYPICAL PROPERTIES OF $ \Sigma^{\infty} $ :
\begin{multline}
 {\mathcal{P}}  ( \Sigma^{\infty} )_{TYPICAL} \; \equiv \; \{ \, p_{\bar{x}}  \in  {\mathcal{P}} (\Sigma^{\infty}) \, :  \\
   \{ \bar{x} \in \Sigma^{\infty} \, : \, p_{\bar{x}} \text{ doesn't hold } \}
   \; \text{is a null set} \}
\end{multline}
\end{slide*}
\begin{slide*}
Denoted by  $ RANDOM( \Sigma^{\infty} ) $ the set of random
sequences over $ \Sigma $  we can restate the
constraint\ref{con:independent sequence of tosses of a classical
coin} as:
\begin{constraint} \label{con:on classical algorithmic randomness}
 \end{constraint}
\textsf{ON THE DEFINITION OF  $ RANDOM( \Sigma^{\infty} )$ }:

 \emph{the unary predicate $ p_{\bar{x}} \; \equiv
\; << \bar{x} \; \in  RANDOM( \Sigma^{\infty} )   >> $ is a
typical property of $\Sigma^{\infty}$, i.e. $ p_{\bar{x}} \in
{\mathcal{P}}  ( \Sigma^{\infty} )_{TYPICAL} $}
\begin{remark}
\end{remark}
Such a constraint doesn't identify  $ RANDOM( \Sigma^{\infty} ) $.
\end{slide*}
\begin{slide*}
It would appear natural to try to characterize the random
sequences over $ \Sigma $ in a purely measure-theoretic way by the
following:
\begin{definition}
\end{definition}
\begin{multline}
  RANDOM( \Sigma^{\infty} )_{purely-measure-theoretic} \; \equiv \\
   \{ \bar{x} \in \Sigma^{\infty} \, : \, p_{\bar{x}} \, holds \, \forall p \in  {\mathcal{P}  ( \Sigma^{\infty} )}_{TYPICAL} \}
\end{multline}
But such a way can't be pursued owing to the following:
\begin{theorem} \label{th:impossibility of absolute conformism}
\end{theorem}
ON THE IMPOSSIBILITY OF ABSOLUTE CONFORMISM:
\begin{equation} \label{eq:impossibility of absolute conformism}
  RANDOM( \Sigma^{\infty} )_{purely-measure-theoretic}  \; = \; \emptyset
\end{equation}
\end{slide*}
\begin{slide*}
\begin{proof}
Following Calude's diagonalization proof \cite{Calude-94} let us
consider the following family of unary predicates over $
\Sigma^{\infty} $ depending on the parameter $ \bar{y} \in
\Sigma^{\infty} $  :
\begin{multline}
   p_{\bar{y}} ( \bar{x} ) \; \equiv \\
    << \, \forall n \in {\mathbb{N}}_{+} \,
\exists m \in {\mathbb{N}}_{+} \; : \\
  m \geq n  \; and \; \bar{x}_{m} \, \neq \, \bar{y}_{m} >>
\end{multline}
Clearly:
\begin{multline}
   P_{unbiased}( \{ \bar{x} \in \Sigma^{\infty} \, : \, p_{\bar{x},  \bar{y} } \text{ doesn't hold} \} ) \; = \;
   0 \\
   \forall \, \bar{y} \in \Sigma^{\infty}
\end{multline}
and so:
\begin{equation}
  p_{\bar{y}} \, \in \,  {\mathcal{P}}  ( \Sigma^{\infty}
  )_{TYPICAL} \; \; \forall \, \bar{y} \in \Sigma^{\infty}
\end{equation}
Anyway:
\begin{equation}
p_{\bar{x}} ( \bar{x} ) \text{ doesn't hold } \; \;  \forall \,
\bar{x} \in \Sigma^{\infty}
\end{equation}
implying the formula eq.\ref{eq:impossibility of absolute
conformism}
\end{proof}
\end{slide*}
\begin{slide*}
\begin{remark}
\end{remark}
CONCEPTUAL DEEPNESS OF MARTIN-L\"{O}F'S RESULT

The theorem\ref{th:impossibility of absolute conformism} shows
that we have to relax the condition that a random sequence
possesses \textbf{all the typical properties} requiring only that
it satisfies \textbf{a proper subclass of typical properties}.

One could , at this point, think that a meaningful restriction
could be obtained again in a purely measure-theoretic framework,
e.g. poning constraints on some kind of speed of convergence to
zero of the unbiased probability of the accepted typical
properties.
\end{slide*}
\begin{slide*}
\textsl{ANYWAY MARTIN-L\"{O}F SHOWED THAT THE RIGHT CRITERIUM OF
SELECTION OF THE PROPER SUBSCLASS DEFINITELY DOESN'T BELONG TO
\textbf{MEASURE THEORY} BUT TO \textbf{COMPUTABILITY THEORY :}}

\textbf{THE CONSIDERED TYPICAL PROPERTIES MUST BE TESTABLE IN AN
\emph{EFFECTIVELY-COMPUTABLE WAY }}
\end{slide*}
\begin{slide*}
\begin{remark}
\end{remark}
MARTIN-L$\ddot{O}$F CONDITION LIES WITHIN THE BOUNDARIES OF
CLASSICAL RECURSION THEORY

By the theorem\ref{th:cardinalities of strings and sequences}:

\begin{itemize}
  \item Computability Theory on $\Sigma^{\star}$ lies within the boundaries of
  \textbf{Classical Recursion Theory} \cite{Odifreddi-89}
  \item Computability Theory on $\Sigma^{\infty}$ lies outside the boundaries of
  \textbf{Classical Recursion Theory}
  \end{itemize}
 Although the definition of a random sequence regards
$\Sigma^{\infty}$ Martin-L\"{o}f's constraint of
effective-computability of the relevant typical properties is
implementable thoroughly in terms of Computability Theory on
$\Sigma^{\star}$ and then belongs to  \textbf{Classical Recursion
Theory} whose firm foundation lies on the theoretic and
experimental evidence lying behind the assumption of
\textbf{Church's Thesis} \cite{Odifreddi-89}, \cite{Odifreddi-96}.
\end{slide*}
\begin{slide*}
\begin{definition}
\end{definition}
$S \; \subset \; \Sigma^{\infty} $ IS ALGORITHMICALLY-OPEN:
\begin{multline}
  ( S \text{ is open } ) \; and \; ( S \, = \, X \Sigma^{\infty} \\
  {\mathbf{X \; recursively-enumerable}})
\end{multline}
\begin{definition}
\end{definition}
ALGORITHMIC SEQUENCE OF ALGORITHMICALLY-OPEN  SETS:

a sequence $ \{ S_{n} \}_{n \geq 1} $ of algorithmically open
sets $  S_{n} \; = \; X_{n} \Sigma^{\infty} $ : $ \exists X \;
\subset \; \Sigma^{\star} \times {\mathbb{N}} $
\textbf{recursively enumerable} with:
\begin{equation*}
   X_{n} \; = \; \{ \vec{x} \in \Sigma^{\star} \, : \, ( \vec{x} , n ) \in
   X \} \; \; \forall n \in {\mathbb{N}}_{+}
\end{equation*}
\end{slide*}
\begin{slide*}
\begin{definition}
\end{definition}
$ S \; \subset \; \Sigma^{\infty} $  IS AN ALGORITHMICALLY-NULL
SET:

$ \exists  \{ G_{n} \}_{n \geq 1} $ algorithmic sequence of
algorithmically-open sets :
\begin{equation*}
  S \; \subset \; \cap_{n \geq 1} G_{n}
\end{equation*}
and:
\begin{equation*}
  alg - \lim_{n \rightarrow \infty} P_{unbiased} ( G_{n} ) \; = \; 0
\end{equation*}
i.e. there exist and increasing, unbounded, \textbf{recursive}
function $ f \, : \, {\mathbb{N}} \rightarrow {\mathbb{N}} $ so
that $ P_{unbiased} ( G_{n} ) \; < \; \frac{1}{2^{k}} $ whenever $
n \; \geq \; f(k) $
\end{slide*}
\begin{slide*}
\begin{definition}
\end{definition}
RANDOM SEQUENCES OVER THE COMMUTATIVE ALPHABET $ \Sigma $ :
\begin{multline}
  RANDOM(\Sigma^{\infty}) \; \equiv \\
   \Sigma^{\infty} - \{ S \subset \Sigma^{\infty} \; \text{ algorithmically null } \}
\end{multline}
\end{slide*}
\begin{slide*}
\section{The difference between commutativity / noncommutativity of the computational device and  commutativity / noncommutativity of the computed objects}
\begin{remark}
\end{remark}
CONFUSION BETWEEN SUBJECT AND OBJECT OF COMPUTATION:

There exists in the literature a partial confusion between the
\textbf{attributes of the computational device} and the
\textbf{attributes of the computed mathematical objects}.
\end{slide*}
\begin{slide*}
Hence some property ( classicality/quantisticality i.e.
commutativity/noncommutativity ) is used in two undistingished (
and often interchanged ) acceptions according to it refers:
\begin{itemize}
  \item to the \textbf{subject of the computation}, i.e. to the
  computational device
  \item to the \textbf{object of the computation}, i.e. to the computed mathematical
  objects
\end{itemize}
\end{slide*}
\begin{slide*}
\begin{remark}
\end{remark}
Any  issue of Computability Theory must analyze separetely each
cell of the following:
\begin{diagram}\label{di:diagram of computation}
\end{diagram}
DIAGRAM OF COMPUTATION:

\begin{tabular}{|c|c|c|c|}
  $ \frac{OBJECT}{SUBJECT} $  & $C_{M}$ & $NC_{M}$  \\
  $C_{\Phi}$  &  $\cdot_{11}$ &  $\cdot_{12}$       \\
  $NC_{\Phi}$ &  $\cdot_{21}$ &  $\cdot_{22}$       \\ \hline
\end{tabular}

with:
\begin{description}
  \item[$C_{M}$ :] MATHEMATICALLY CLASSICAL
  \item[$NC_{M}$:] MATHEMATICALLY NONCLASSICAL
  \item[$C_{\Phi}$:] PHYSICALLY CLASSICAL
  \item[$NC_{\Phi}$:] PHYSICALLY NONCLASSICAL
\end{description}
\end{slide*}
\begin{slide*}
\textmd{$ 1^{th}$ ISSUE: WHO IS COMPUTABLE ?}

\begin{itemize}
  \item $cell_{11} \; : \; C_{M} \, \cap \, C_{\Phi} $

  There is complete agreement in the scientific community that,
  as to the computation by \textbf{physically classical
  computers} of the following set of functions:
\begin{definition}
\end{definition}
MATHEMATICALLY CLASSICAL FUNCTIONS:

    (partial) functions on sets  $ S \, : \, card(S) \, \leq \,
  \aleph_{0}$

\textbf{Church's Thesis} holds leading to the identificaton of the
computable (partial) functions with the (partial) recursive
functions \cite{Odifreddi-89}, \cite{Odifreddi-96}
\end{itemize}
\end{slide*}
\begin{slide*}
\begin{itemize}
  \item $cell_{21} \; : \; C_{M} \, \cap \, NC_{\Phi} $

There is no universally accepted  answer in the scientific
community to the question if a \textbf{physically nonclassical
computer} can violate Church's  Thesis, i.e. can computate
non-recursive \textbf{mathematically classical functions}.

In particular, as far as  the computation by \textbf{physically
quantistical computers} of \textbf{mathematically classical
functions} is concerned, the common opinion among the leading
researchers in Quantum Computation \cite{Feynman-82},
\cite{Deutsch-85}, \cite{Jozsa-98} is that \textbf{Nonrelativistic
Quantum Mechanics} and \textbf{Partially-relativistic Quantum
Mechanics (Local Quantum Field Theories)} don't violate Church's
Thesis.
\end{itemize}
\end{slide*}
\begin{slide*}
Finally, when \textbf{Generally-relativistic Quantum Mechanics}
(both in the form of \textbf{quantum Gravity} and in the form of
some suggested \textbf{gravitationally-modificated Quantum
Mechanics}) is considered, the whole story touches the strongly
debated ideas of R. Penrose \cite{Penrose-89}, \cite{Penrose-96}
\end{slide*}
\begin{slide*}
\begin{itemize}
  \item $cell_{12} \; : \; NC_{M} \, \cap \, C_{\Phi} $

  As soon as one goes out from the boundaries of Classical
  Recursion Theory the almost miracolous equivalence of  all the
  different approaches, that in such a  theory manifests the strong
  experimental verification of Church's Thesis, dramatically
  disappears.
\end{itemize}
\end{slide*}
\begin{slide*}
   Just as to the Computability Theory by \textbf{physically classical
  computers} of  (partial) functions on sets  $ S \, : \, card(S) \, = \,
  \aleph_{1}$  many different inequivalent candidate theories have
  been proposed:
\begin{enumerate}
  \item the Standard Theory generated by the studies of Grzegorczyck -
  Lacombe \cite{Pour-El-Richards-89}
  \item the theory developed by the so called Markov School in the framework of
  Constructive Mathematics \cite{Odifreddi-89}
  \item the Blum - Shub - Smale 's
  Theory \cite{Smale-92}, \cite{Blum-Cucker-Shub-Smale-98}
\end{enumerate}
The relative popularity of the issue about the concurrence of
such candidate theories is owed to Penrose's question if
Mandelbrot set is recursive \cite{Penrose-89}.
\end{slide*}
\begin{slide*}
Given a \textbf{noncommutative probability space} $ ( A , \omega
) $:
\begin{definition}
\end{definition}
AUTOMORPHISMS OF A:
\begin{equation}
  Aut(A) \; \equiv \; \{ \alpha \, : \, \text{ involutive morphisms of
  A } \}
\end{equation}
\begin{definition}
\end{definition}
DYNAMICS OF $( A , \omega )$ \cite{Benatti-93}:
\begin{multline}
  DYN[( A , \omega )] \; \equiv \; \{ \alpha \in Aut(A) : \\
   \omega ( \alpha ( a )) \, = \, \omega ( a ) \; \forall a \in A \}
\end{multline}
\end{slide*}
\begin{slide*}
\begin{definition}
\end{definition}
$ C_{\phi} $ - COMPUTABLE AUTOMORPHISMS OF A:
\begin{multline}
 C_{\phi} - AUT( A ) \; \equiv \\
  \{ \alpha \in AUT( A ) \, : \\
  \alpha \text{ is computable by $classical_{\Phi}$ computers} \}
\end{multline}
\begin{definition}
\end{definition}
$ C_{\phi} $ - COMPUTABLE-DYNAMICS OF $( A , \omega )$:
\begin{multline}
 C_{\phi} - DYN[( A , \omega )] \; \equiv \\
  \{ \alpha \in DYN[( A , \omega )] \, : \\
  \alpha \text{ is computable by $classical_{\Phi}$ computers} \}
\end{multline}
\end{slide*}
\begin{slide*}
\begin{itemize}
  \item $cell_{22} \; : \; NC_{M} \, \cap \, NC_{\Phi} $

  It's important to realize that Church Thesis doesn't imply that the answer to the $ 1^{th} ISSUE $ contained in the cells
  $cell_{12}$ and $cell_{22}$ must be equal.

  For example Church Thesis is not incompatible with an hypothetical situation in which  Mandelbrot
  set would be $ C_{\Phi} $ - incomputable but $ NC_{\Phi} $ - computable
\end{itemize}
\end{slide*}
\begin{slide*}
In the same way , given a \textbf{noncommutative probability
space} $ ( A , \omega ) $ and introduced the following notions:
\begin{definition}
\end{definition}
$ NC_{\phi} $ - COMPUTABLE AUTOMORPHISMS OF A:
\begin{multline}
 NC_{\phi} - AUT( A ) \; \equiv \\
  \{ \alpha \in AUT( A ) \, : \\
  \alpha \text{ is computable by $nonclassical_{\Phi}$ computers} \}
\end{multline}
\begin{definition}
\end{definition}
$ NC_{\phi} $ - COMPUTABLE-DYNAMICS OF $( A , \omega )$:
\begin{multline}
 NC_{\phi} - DYN[( A , \omega )] \; \equiv \;
  \{ \alpha \in DYN[( A , \omega )] \, : \\
  \alpha \text{ is computable by
   $nonclassical_{\Phi}$ computers} \}
\end{multline}
\end{slide*}
\begin{slide*}
we have that:
\begin{multline}
   \text{Church Thesis}  \; \nRightarrow \\
( C_{\phi} -  AUT( A ) \, = \, NC_{\phi} AUT( A ))
\end{multline}
\begin{multline}
   \text{Church Thesis}  \; \nRightarrow \\
( C_{\phi} -  DYN[( A , \omega
  )] \, = \, NC_{\phi} - DYN[( A , \omega
  )] )
\end{multline}
\end{slide*}
\begin{slide*}
\textmd{$ 2^{th}$ ISSUE: WHO IS EFFICENTELY COMPUTABLE ?}

The deep scientific revolution brought by Quantum Computation is
that:

\emph{\textbf{Computational Complexity Theory is not a purely
mathematical theory \cite{Odifreddi-99} in that the answers it
gives are different on the $ 1^{th}$ and the $ 2^{th}$ rows of the
diagram\ref{di:diagram of computation}}}

as  is ultimatively implied by the complexity class relations
\cite{Bernstein-Vazirani-97}, \cite{Williams-Clearwater-98}:
\begin{equation} \label{eq:first quantum complexity classes relation}
  P \;  \subset \; QP
\end{equation}
\begin{equation} \label{eq:second quantum complexity classes relation}
  ZPP \;  \subset \; ZQP
\end{equation}
\end{slide*}
\begin{slide*}
\begin{remark}
\end{remark}
\textbf{QUANTUM DICE} DIFFERS BOTH FROM \textbf{CLASSICAL DICE}
AND FROM \textbf{CLASSICAL ANA${\mathbf{\Gamma}}$KH}

The relations eq.\ref{eq:first quantum complexity classes
relation}, eq.\ref{eq:second quantum complexity classes relation}
show that deep peculiarity of the statistical structure of Quantum
Mechanics \cite{Holevo-99}:

they ultimatively imply that, under the assumption $ P \; \neq \;
NP $ \cite{Odifreddi-99}, \textbf{quantum nondeterminism} is
different both from \textbf{classical determinism} and from
\textbf{classical nondeterminism}.

Unfortunately such an issue has not been considered yet in all the
discussions about the possibility of a deterministic completion of
Quantum Mechanics \cite{Wheeler-Zurek-83}, \cite{Bell-93},
\cite{Peres-95}, \cite{Bohm-Hiley-93}, \cite{Svozil-98},
\cite{Auletta-00}
\end{slide*}
\begin{slide*}
\textsl{\emph{\textbf{FUNDAMENTAL QUESTION :}}}

\textsl{\emph{\textbf{DOES ALGORITHMIC INFORMATION THEORY DIFFERS
IN THE $ 1^{TH}$ AND IN THE $ 2^{TH}$ ROWS OF THE
DIAGRAM\ref{di:diagram of computation} ?}}}
\end{slide*}
\begin{slide*}
\begin{remark}
\end{remark}
ARGUMENT TO ANSWER $<<YES>>$ TO THE \textbf{FUNDAMENTAL QUESTION}:

By the link existing between \textbf{Computational Complexity
Theory} and \textbf{Algorithmic Information Theory} ( passing,
mainly, through \textbf{resource-bounded algorithmic information }
\cite{Longpre-92}, \cite{Calude-94}, \cite{Li-Vitanyi-97} ) and
the relations eq.\ref{eq:first quantum complexity classes
relation}, eq.\ref{eq:second quantum complexity classes relation}
\end{slide*}
\begin{slide*}
\section{Quantum Algorithmic Information Theory and the Pour El extension of Church Thesis}
\begin{remark}
\end{remark}
ARGUMENT TO ANSWER $<<NO>>$ TO THE \textbf{FUNDAMENTAL QUESTION}:

If one assumed that:
\begin{enumerate}
  \item \textbf{Quantum Algorithmic Information Theory} must satisfy \textbf{Uspensky's
Axiomatic Construction} \cite{Uspensky-92}
  \item \textbf{Pour El Thesis} \cite{Pour-El-99} holds
\end{enumerate}

it would  follow that for finite dimensional quantum systems the
answer to the fundamental question is $ << no >> $.
\end{slide*}
\begin{slide*}
Algorithmic Information Theory , i.e. the theory dealing with the
algorithmic information of an object defined as  the length of the
shortest algorithm calculating
 it, has been originally defined for sets of objects with
 cardinality at most $ \aleph_{0} $ \cite{Calude-94}.

A generalization of such a theory have been proposed by Vladimir
 A. Uspensky through the introduction of an axiomatic procedure by
 which Algorithmic Information Theory may be contructed on any
 set of objects satisfying certain properties.

 Demanding to the original Uspensky's article \cite{Uspensky-92}
 for details I will briefly review here what I will call from now on Uspensky's Axiomatic
 Procedure.
\end{slide*}
\begin{slide*}
Given a  set S let us introduce the following definitions:
\begin{definition} \label{def:length on a set}
\end{definition}
LENGTH ON S :
\begin{equation}\label{eq:length on a set}
l : S \rightarrow {\mathbb{R}}_{+} \cup \{0 \}
\end{equation}

\begin{definition} \label{def:lengthed set}
\end{definition}
LENGTHED SET:
\begin{equation}\label{eq:lengthed set}
    \text{a couple} \, ( S \, , \, l \, ) :  S \, \text{is a set and l is a length on S}
\end{equation}
\end{slide*}
\begin{slide*}
Given a set $ S $  let us define:
\begin{definition} \label{def:partial functions}
\end{definition}
SET OF THE PARTIAL FUNCTIONS ON S :
\begin{equation} \label{eq:partial functions}
  PF ( S ) \; \equiv \; \{ \phi : S \stackrel{\circ}{\rightarrow}
  S \}
\end{equation}
\end{slide*}
\begin{slide*}
Given a lengthed set $ ( S \, , \, l \, ) $ let us define:
\begin{definition} \label{def:descriptive information}
\end{definition}
DESCRIPTIVE INFORMATION ON $ ( S \, , \, l \, ) $ W.R.T. $ \phi
\in PF ( S ) $ :

 $ I_{\phi} : S \rightarrow {\mathbb{R}}_{+} \cup \{0 , \infty \} $ :
\begin{equation}
  I_{\phi} (  y ) \equiv
  \begin{cases}
    min  \{ l(x)  :  \phi ( x ) =  y  \}  &   \exists x \in S :  \phi ( x ) = y   \\
    + \infty & \text{otherwise}.
  \end{cases}
\end{equation}
\end{slide*}
\begin{slide*}
Given , then, a set $ {\mathcal{C}} \, \subseteq PF(S) $ we can
introduce on it the following partial ordering:
\begin{definition}
\end{definition}
$ \phi_{1} \in {\mathcal{C}} $ IS LESS PROLIX THAN $ \phi_{2} \in
{\mathcal{C}} \; (   \phi_{1} \, \leq \, \phi_{2} ) $ :
\begin{equation}
  \exists \, c_{\phi_{1} , \phi_{2} } \in  {\mathbb{R}}_{+} \; :
  I_{\phi_{1}} (x) \, \leq \,  I_{\phi_{2}} (x) + c_{\phi_{1} , \phi_{2} } \; \forall x \in S
\end{equation}

We will say, then, that:
\begin{definition}
\end{definition}
$ \phi_{1}  \in {\mathcal{C}} $ AND $ \phi_{2}  \in {\mathcal{C}}
$ ARE EQUIVALENT  $ (   \phi_{1} \, \sim \, \phi_{2} ) $ :
\begin{equation}
  ( \, \phi_{1}  \, \leq \, \phi_{2} \, ) \; and \; ( \, \phi_{2}  \, \leq \, \phi_{1} \, )
\end{equation}
\end{slide*}
\begin{slide*}
Let us now introduce the following basic notions:
\begin{definition}
\end{definition}
OPTIMAL DESCRIPTIVE METHOD  IN $ {\mathcal{C}} $ :
\begin{equation}
  \omega \, \in \, min_{\leq} \, \frac{{\mathcal{C}}}{\sim}
\end{equation}
\begin{definition} \label{def:objectivity}
\end{definition}
DESCRIPTIVE INFORMATION BY $ {\mathcal{C}} $ IS OBJECTIVE:
\begin{equation} \label{eq:objectivity}
  \exists \, min_{\leq} \, \frac{{\mathcal{C}}}{\sim}
\end{equation}
\end{slide*}
\begin{slide*}
\begin{remark}
\end{remark}
PASSAGE FROM DESCRIPTIVE INFORMATION TO ALGORITHMIC INFORMATION:

Let us observe that, up to now, I have spoken about
\emph{descriptive information} and not of \emph{algorithmic
information}: in fact I have not yet introduced the more
important constraint on the allowed description methods: that of
being \emph{algorithmically implementable}, or, said in a
different way, to be \emph{effectively-computable w.r.t. the
informal notion of effective-computability}.

Though such a passage was proposed by A.N. Kolmogorov to bypass
the problem that \textbf{descriptive information by $PF(
\Sigma^{\star})$ was not objective} the conceptual meaning of
resorting to \textbf{Computability Theory} was extraordinarily
clear to the great mathematician \cite{Shiryayev-94}.
\end{slide*}
\begin{slide*}
\begin{definition}
\end{definition}
$ C_{\Phi}$ - COMPUTABLE-PARTIAL FUNCTIONS ON S:
\begin{multline}
   C_{\Phi} - PF(S) \; \equiv \\
   \{ f \in PF(S)\, : \, f \text{ is computable} \\
    \text{by $classical_{\Phi}$ computers} \}
\end{multline}
\begin{definition}
\end{definition}
$ NC_{\Phi}$ - COMPUTABLE-PARTIAL FUNCTIONS ON S:
\begin{multline}
   NC_{\Phi} - PF(S) \; \equiv \\
   \{ f \in PF(S)\, : \, f \text{ is computable} \\
   \text{by $nonclassical_{\Phi}$ computers} \}
\end{multline}
\end{slide*}
\begin{slide*}
We have now all the ingredients required to completely formalize
the Uspensky's Axiomatic Procedure:

\textbf{USPENSKY'S AXIOMATIC PROCEDURE TO INTRODUCE
PHYSICALLY-CLASSICAL AND PHYSICALLY-NONCLASSICAL ALGORITHMIC
INFORMATION THEORY} ON A LENGTHED SET $ ( S \, , \, l \, ) $:

\begin{itemize}
  \item $ C_{\Phi} \; ( NC_{\Phi} ) $- ALGORITHMIC INFORMATION THEORY ON  $ ( S \, , \, l \, ) $ MAY BE DEFINED IF AND
  ONLY IF  DESCRIPTIVE INFORMATION ON
$C_{\Phi}-PF(S) \; ( NC_{\Phi}-PF(S) ) $ IS OBJECTIVE
\end{itemize}
\end{slide*}
\begin{slide*}
\begin{itemize}
  \item THE $ C_{\Phi} \; ( NC_{\Phi} ) $- ALGORITHMIC INFORMATION THEORY ON $ ( S \, , \, l \, ) $ IS DEFINED AS THE DESCRIPTIVE
  INFORMATION W.R.T. AN OPTIMAL DESCRIPTIVE METHOD  IN  A CERTAIN SUBSET:
\begin{align*}
  C_{\Phi}&-AC-AL( S )  \; \subseteq
  C_{\Phi}-PF(S) \\
  ( NC_{\Phi}&-AC-AL( S  )  \; \subseteq
  NC_{\Phi}-PF(S ))
\end{align*}
  \end{itemize}
\end{slide*}
\begin{slide*}
\begin{remark}
\end{remark}
EXTENSION OF THE ABOVE CONSTRUCTION TO STRUCTURED SETS:

Eventually  S might be endowed with some suppletive structure $
\circledS $. The objects we want to describe will , then, be
considerated , more properly, as elements  of the mathematical
structure ( S , l , $ \circledS $ ).

Our descriptional process will, then, have to take in
consideration such a structure. The considered class  of
description-methods shall, than , consist of subsets not of PF(S)
but of its subset:
\begin{definition}
\end{definition}
SET OF THE PARTIAL ISOMORPHISMS OF

 ( S , $ \circledS $ ):
\begin{equation}
  PI ( \, S \, , \, \circledS \, ) \; \equiv \; \{ f \in PF(S) \,
  : \, \text{ f is $ \circledS $ - preserving} \}
\end{equation}
\end{slide*}
\begin{slide*}
\begin{definition}
\end{definition}
$ C_{\Phi}$ - COMPUTABLE-PARTIAL ISOMORPHISMS ON ( S , $ \circledS
$ ):
\begin{multline}
   C_{\Phi} - PI(S,  \circledS ) \; \equiv \\
   \{ f \in C_{\Phi} - PI(S) \, : \\
    f \text{ is computable} \\
    \text{by $classical_{\Phi}$ computers} \}
\end{multline}
\begin{definition}
\end{definition}
$ NC_{\Phi}$ - COMPUTABLE-PARTIAL ISOMORPHISMS ON ( S , $
\circledS $ ):
\begin{multline}
   NC_{\Phi} - PI(S,  \circledS ) \; \equiv \\
   \{ f \in NC_{\Phi} - PI(S) \, : \\
   \, f \text{ is computable} \\
    \text{by $nonclassical_{\Phi}$ computers} \}
\end{multline}
\end{slide*}
\begin{slide*}
\textbf{USPENSKY'S AXIOMATIC PROCEDURE TO INTRODUCE
PHYSICALLY-CLASSICAL AND PHYSICALLY-NONCLASSICAL ALGORITHMIC
INFORMATION THEORY} ON A  STRUCTURED LENGTHED SET ( S , l , $
\circledS $ )

\begin{itemize}
  \item $ C_{\Phi} \; ( NC_{\Phi} ) $- ALGORITHMIC INFORMATION THEORY ON  ( S , l , $
\circledS $ ) MAY BE DEFINED IF AND
  ONLY IF  DESCRIPTIVE INFORMATION ON
$C_{\Phi}-PI( S ,   \circledS  ) \; ( NC_{\Phi}-PI( S , \circledS
) ) $ IS OBJECTIVE
\end{itemize}
\end{slide*}
\begin{slide*}
\begin{itemize}
  \item THE $ C_{\Phi} \; ( NC_{\Phi} ) $- ALGORITHMIC INFORMATION THEORY ON ( S , l , $
\circledS $ ) IS DEFINED AS THE DESCRIPTIVE
  INFORMATION W.R.T. AN OPTIMAL DESCRIPTIVE METHOD  IN  A CERTAIN
  SUBSET:
\begin{align*}
  C_{\Phi}&-AC-AL( S ,\circledS  )  \; \subseteq
  C_{\Phi}-PI(PS ,\circledS  ) \\
  ( NC_{\Phi}&-AC-AL( S ,\circledS  )  \; \subseteq
  NC_{\Phi}-PI(PS ,\circledS  ))
\end{align*}
\end{itemize}
\end{slide*}
\begin{slide*}
Marian Boykan Pour-El and Jonathan Ian Richards has developed a
very interesting Computability Theory on Banach Spaces
\cite{Pour-El-Richards-89}  that, under the explicit assumption of
a generalization of \textbf{Church Thesis} that I will call from
now on \textbf{Pour El Thesis} \cite{Pour-El-99} characterizes
mathematically:
\begin{enumerate}
  \item a subset:
\begin{equation*}
  B_{COMP} \; = \; C_{\Phi}- B \; = \;  NC_{\Phi}- B
\end{equation*}
of \textbf{vectors} of a  \textbf{Banach space} B
  \item a subset:
  \begin{equation*}
  C_{\Phi} - {\mathcal{L}}({\mathbb{H}}) \; = \; NC_{\Phi} - {\mathcal{L}}({\mathbb{H}})
  \; \subset \; {\mathcal{L}}({\mathbb{H}})
\end{equation*}
of the space $ {\mathcal{L}}({\mathbb{H}}) $ of the
\textbf{linear operators} on a \textbf{separable Hilbert space} $
{\mathbb{H}}$
\end{enumerate}
that are \textbf{effectively computable, according to the
informal notion of effective computability, by any kind of
physical computer ( classical or nonclassical )}
\end{slide*}
\begin{slide*}
Given a Banach space B  on the real/complex field Pour-El and
 Richards introduce the following notion:
 \begin{definition} \label{def:computability structure on a Banach space}
 \end{definition}
 COMPUTABILITY STRUCTURE ON B:

 a specification of a subset $ {\mathcal{S}} $ of the set $ B^{\infty} $ of all
 the sequences  in B identified as the \textbf{set of the
 computable sequences on B} satisfying the following axioms:
\end{slide*}
\begin{slide*}
\begin{axiom}  \label{ax:linear forms}
\end{axiom}
ON LINEAR FORMS:

 \begin{hypothesis}
 $ \{ x_{n} \} $ and  $ \{ y_{n} \} $ computable sequences in B

 $ \{ \alpha_{n,k} \} ,  \{ \beta_{n,k} \} $ two recursive double
 sequence of real/complex numbers

 d recursive function

 $ s_{n} \equiv \sum_{k=0}^{d(n)} \alpha_{n,k} x_{k} + \beta_{n,k}
 y_{k} $
 \end{hypothesis}
 \begin{thesis}
 $ \{ s_{n} \} \in { \mathcal{S} } $
 \end{thesis}
\end{slide*}
\begin{slide*}
\begin{axiom} \label{ax:limits}
\end{axiom}
ON LIMITS:

 \begin{hypothesis}
 $ x_{n,k} $ computable double sequence in B : $ alg- \lim_{k
 \rightarrow \infty} x_{n,k} =  x_{n}  $
 \end{hypothesis}
 \begin{thesis}
 $ \{ x_{n} \} \in {\mathcal{S}} $
 \end{thesis}
\end{slide*}
\begin{slide*}
\begin{axiom}  \label{ax:norms}
\end{axiom}
ON NORMS:

 \begin{hypothesis}
 $ \{ x_{n} \} \in {\mathcal{S}} $
 \end{hypothesis}
 \begin{thesis}
 $ \{ \| x_{n} \| \} $ is a  recursive sequence of real  numbers.
 \end{thesis}
where:
\end{slide*}
\begin{slide*}
\begin{definition}
\end{definition}
THE SEQUENCE OF RATIONAL NUMBERS $ \{ r_{n} \} $ IS COMPUTABLE:

$ \exists \; a,b,c $ recursive functions:
\begin{equation}
\begin{split}
  ( c_{n} \; \neq & \; 0 \; \forall n) \; and \\
  ( r_{n} \, = & \, (-1)^{a(n)} \, \frac{b(n)}{c(n)} )
\end{split}
\end{equation}
THE SEQUENCE OF RATIONAL NUMBERS $ \{ r_{n} \} $ CONVERGES
ALGORITHMICALLY TO $ x \in {\mathbb{R}}$  ( $ alg - \lim_{n
\rightarrow \infty}  r_{n} \; = \; x$ )
\begin{equation}
\begin{split}
   \exists f & \text{recursive function} :  \\
  n \geq f(n) \; & \Rightarrow \; | r_{n} - x | < \frac{1}{2^{n}}
\end{split}
\end{equation}
\begin{definition}
\end{definition}
RECURSIVE REAL NUMBERS:
\begin{multline}
  {\mathbb{R}}_{COMP} \; \equiv \\
   \{ x \in {\mathbb{R}} \, : \exists  \{ r_{n} \} \text{ computable sequence of rationals} : \\
     alg - \lim_{n \rightarrow \infty} r_{n} \; = \; x  \}
\end{multline}
\end{slide*}
\begin{slide*}
SOME PROPERTIES OF ${\mathbb{R}}_{COMP}$:
\begin{enumerate}
  \item $ ( \, {\mathbb{R}}_{COMP} \, , \, + \, , \, \cdot \, ) $ is a field
  \item $ \pi \, , \, e \, , \, \gamma \; \in \; {\mathbb{R}}_{COMP} $
  \item
\begin{equation}
  {\mathbb{R}}_{ALGEBRAIC} \; \subset \; {\mathbb{R}}_{COMP}
\end{equation}
   \item
\begin{equation}
  card(  {\mathbb{R}}_{COMP} ) \; = \; \aleph_{0}
\end{equation}
\end{enumerate}
\end{slide*}
\begin{slide*}
Given a double sequence of real numbers  $ \{ x_{n,k} \ \} $ and
an other sequence  $ \{ x_{n} \} $ of real numbers such that:
\begin{equation}
 \lim_{ k \rightarrow \infty }  x_{n,k} = x_{n} \;  \forall n \in
 \mathbb{N}
\end{equation}
\begin{definition}
\end{definition}
 $ \{ x_{n,k} \} $ CONVERGES ALGORITHMICALLY TO $ \{ x_{n} \} ( alg- \lim_{k \rightarrow \infty } x_{n,k} = x_{n}
 ) $
 \begin{multline}
 \exists e: { \mathbb{N} } \times { \mathbb{N} } \rightarrow {
 \mathbb{N} } \; recursive  \; : \\
  ( k > e(n,N) \Rightarrow \mid
  x_{n,k} - x_{n} \mid \leq  \frac{1}{2^{N}} ) \; \forall n \in { \mathbb{N}
 } , \; \forall N \in { \mathbb{N} }
\end{multline}
\begin{definition}
\end{definition}
$ \{ x_{n} \}_{n \in {\mathbb{N}}} $ IS COMPUTABLE:
 \begin{multline}
 \exists \{ r_{n,k} \in { \mathbb{Q} } \}_{n,k \in { \mathbb{N} }
 } \; computable \; : \\
  \mid r_{n,k} - x_{n} \mid \leq \frac{1}{2^{k}} \; \; \forall n,k
  \in { \mathbb{N} }
 \end{multline}
\end{slide*}
\begin{slide*}
\begin{remark}\label{rem:computability of sequnce as starting point}
\end{remark}
THE COMPUTABILITY OF A SEQUENCE IS MORE THAN THE COMPUTABILITY OF
ALL ITS ELEMENTS

 given a sequence  $ \{ x_{n} \}
 $ of real numbers, the fact that each element  of the
 sequence is computable, and can, consequentely, be effectively
 approximated to any desired degree of  precision  by a computer
 program $ P_{n} $ given in advance doesn't imply the
 computability of the whole sequence since there might not exist an erffective way of
 combining the sequence of programs $ \{ P_{n} \} $  in a unique
 program P computing the whole sequence $ \{ x_{n} \} $.
\end{slide*}
\begin{slide*}
 Remark\ref{rem:computability of sequnce as starting point} should clarify why the definition of a
 computability structure on a Banach space B is made through a
 proper specification of  the computable sequences in B and not,
 simply, by  the specification of a proper set of  the computables
 vectors.

 The notion of a computable vector, instead, is immediately induced
 by the assignment on  B of a computability structure $ \mathcal{S}
 $.
\begin{definition}
\end{definition}
COMPUTABLE VECTORS OF B:
 \begin{equation}
 B_{COMP} \equiv \{ x \in B : \{x,x,x, \ldots \} \in
 \mathcal{S} \}
 \end{equation}
\end{slide*}
\begin{slide*}
\begin{remark}
\end{remark}
INTUITIVE MEANING OF THE AXIOMS Axiom\ref{ax:linear forms},
Axiom\ref{ax:limits} and Axiom\ref{ax:norms}

since a Banach space is  made up of:
\begin{enumerate}
  \item a linear space V
  \item a norm on V
  \item the completeness-condition for such a norm
\end{enumerate}
it appears natural to require analogous effective conditions for
the set of computable sequences.
\end{slide*}
\begin{slide*}
\begin{remark}
\end{remark}
THE MULTIVOCITY PROBLEM FOR THE COMPUTABILITY STRUCTURE

The axioms Axiom\ref{ax:linear forms}, Axiom\ref{ax:limits} and
Axiom\ref{ax:norms} don't provide the axiomatic definition of a
unique structure for a Banach  space B  since B admits,
generally, more computability-structures.

This, anyway, doesn't relativize the whole approach thanks to the
existence of a suppletive condition whose satisfability results in
the invoked univocity.
\end{slide*}
\begin{slide*}
Given a computability structure  $ \mathcal{S} $  on a Banach
space B:
\begin{definition}
\end{definition}
EFFECTIVE GENERATING SET FOR B:
 \begin{multline}
  \{ e_{n} \} \in {\mathcal{S}} \; : \\
   linear-span( \{ e_{n} \} ) \; is \; dense\; in \; B
 \end{multline}
\begin{definition}
\end{definition}
B IS EFFECTIVELY SEPARABLE:
\begin{equation}
 \exists \{ e_{n} \} \; effective \; generating \; set \; for \; B
\end{equation}
\end{slide*}
\begin{slide*}
\begin{theorem}\label{th:of univocity}
\end{theorem}
THEOREM OF UNIVOCITY

 \begin{hypothesis}
 B Banach space

 $ {\mathcal{S}}_{1} $ , $ {\mathcal{S}}_{2} $  effectively separable computability structures on B

 $ \{ e_{n} \} \in {\mathcal{S}}_{1} \cap {\mathcal{S}}_{2} $
 effective generating set for B
 \end{hypothesis}
 \begin{thesis}
 $ {\mathcal{S}}_{1} = {\mathcal{S}}_{2} $
 \end{thesis}
\end{slide*}
\begin{slide*}
\begin{remark}
\end{remark}
COMPUTABILITY STRUCTURE OF A QUANTUM SYSTEM:

Given a \textbf{quantum physical  system }$ ( \, {\mathcal{H}} \,
, \, \hat{H} \, ) $ the existence of an effectively measurable
operator having as eigenvectors a basis  $ \{ e_{n} \} $ of $
\mathbb{H} $ gives us immediately an univocal notion of
computability on $ \mathbb{H} $: that associated to the effective
generating set $ \{ e_{n} \} $ (said an \textbf{effective-basis}
of $ \mathbb{H} $).
\end{slide*}
\begin{slide*}
\begin{example} \label{ex:one half spin system}
\end{example}
SPIN $\frac{1}{2}$ SYSTEMS

Given a \textbf{quantum physical  system } $ ( \, {\mathcal{H}}
\; =  \;{\mathbb{C}}^{2} \; , \; \hat{H} \; = \; f (
\hat{\sigma_{x}} , \hat{\sigma_{y}} , \hat{\sigma_{z}} ) ) $
since the x-component, the y-component and the z-component  of the
spin are observable effectively-measurable (e.g. by a
Stern-Gerlach apparatus)  it follows  that :
\begin{equation*}
\begin{split}
  \{ &
\begin{pmatrix}
  1 \\
  0
\end{pmatrix}   \, , \, \begin{pmatrix}
  0 \\
  1
\end{pmatrix} \, \} \, , \, \{
\begin{pmatrix}
  \frac{1}{\sqrt{2}} \\
  \frac{1}{\sqrt{2}}
\end{pmatrix}   \, , \, \begin{pmatrix}
   \frac{1}{\sqrt{2}} \\
  - \frac{1}{\sqrt{2}}
\end{pmatrix} \, \}  \\
  \{ &
\begin{pmatrix}
  \frac{1}{\sqrt{2}} \\
  \frac{i}{\sqrt{2}}
\end{pmatrix}   \, , \, \begin{pmatrix}
   \frac{1}{\sqrt{2}} \\
  - \frac{i}{\sqrt{2}}
\end{pmatrix}  \, \}
\end{split}
\end{equation*}
are three \textbf{effective-bases} of $ {\mathbb{H}} $.
\end{slide*}
\begin{slide*}
Furthermore since also the identity operator is obviously
effectively measurable it follows that $ \{ \, {\mathbb{I}} \, ,
\, \sigma_{x} \, , \, \sigma_{y} \, , \,\sigma_{z} \, \} $ is an
\textbf{effectively generating set} for the $W^{\star}$-algebra
$  {\mathcal{B}}({\mathbb{H}}) \; = \; M_{2}({\mathbb{C}}) $.
\end{slide*}
\begin{slide*}
Given  an \textbf{effectively separable Hilbert space} $ {
\mathbb{H}} $
\begin{definition} \label{def:computable operator}
\end{definition}
COMPUTABLE LINEAR OPERATOR ON  $ { \mathbb{H}} \; ( T \in
{\mathcal{L}}_{COMP} ({\mathbb{H} } ) )$

$ T \in {\mathcal{L}} ({\mathbb{H} })$ closed, such that there
exist a computable sequence $ \{e_{n}\} $ in $ {\mathbb{H}}$ so
that:

\begin{multline}
\{ ( e_{n} , T, e_{n} ) \} \text{ is a  computable sequence} \\
of \; {\mathbb{H}} \times {\mathbb{H}}
\end{multline}
and:
\begin{multline}
linear-span \{ ( e_{n} , T, e_{n} ) \} \text{ is dense in} \\
\text{the graph }  \Gamma (T) \; of \; T
\end{multline}
\end{slide*}
\begin{slide*}
\begin{remark}
\end{remark}
INTUITIVE MEANING OF THE DEFINITION \ref{def:computable operator}
\begin{itemize}
  \item a \textbf{bounded operator} is computable if its action on any computable vector is effectively
determinable
  \item an \textbf{unbounded operator}  is computable if its action on any computable vector is effectively
determinable and if we are able to solve effectively the \textbf{
halting problem} corresponding to the belongness to its domain of
definition, i.e. if we have an effective-algorithm that , given a
generic computable vector x of $ {\mathbb{H}} $ tells us whether
T halts  on x $ (Tx \downarrow ) $ or not $ (Tx \uparrow ) $.
\end{itemize}
\end{slide*}
\begin{slide*}
\begin{remark}
\end{remark}
FACTORS AS BUILDING BLOCKS OF VON NEUMANN ALGEBRAS:

Any $ W^{\star} $-algebra A is a sort of direct integral of
factors:
\begin{equation} \label{eq:factor decomposition}
  A \; = \; \int_{{\mathcal{Z}}(A)}^{\otimes} A_{\lambda} \, d \nu ( \lambda )
\end{equation}
where:
\begin{itemize}
  \item  $ {\mathcal{Z}}(A)  \; \equiv \; A \cap A' $ is the
  \textbf{center} of A
  \item the $ A_{\lambda} $ are all \textbf{factors}, i.e.:
\begin{equation}
  {\mathcal{Z}}(A_{\lambda}) \; = \; \{ {\mathbb{C}} \,
  {\mathbb{I}} \}
  \; \; \forall \lambda \in  {\mathcal{Z}}(A)
\end{equation}
\end{itemize}
Hence the analysis of a $ W^{\star} $-algebra may be reduced to
the analysis of its building blocks
\end{slide*}
\begin{slide*}
\begin{definition}
\end{definition}
DISCRETE TYPE VON NEUMANN ALGEBRA:

a $ W^{\star} $ -algebra in which \textbf{factor decomposition}
eq.\ref{eq:factor decomposition} appear only factors of type $
I_{n} \, n \in {\mathbb{N}} \cup \{ \infty \} $ , i.e. don't
appear factors of type $ II_{n} \; n \in \{ 1 , \infty \}$ and of
type $ III_{\alpha} \; \alpha \in  [ 0 , 1 ] $
\begin{definition}
\end{definition}
DISCRETE TYPE NONCOMMUTATIVE PROBABILITY SPACE:

$ ( \, A \, , \, \omega \, ) $ noncommutative probability space
with A discrete type $W^{\star}$-algebra
\end{slide*}
\begin{slide*}
\begin{remark}
\end{remark}
POUR EL THESIS TOUCHES ONLY DISCRETE TYPE NONCOMMUTATIVE
PROBABILITY SPACES

Since a $W^{\star}$-algebra is isomorphic to the space $
{\mathcal{B}} ({\mathbb{H}}) $ of the \textbf{bounded linear
operators on a separable Hilbert space} $ {\mathbb{H}} $ if and
only if it is of \textbf{discrete type} \cite{Benatti-93} it
follows that Pour El Thesis implies the following relations:
\begin{multline}
  C_{\Phi} - AUT (A) \; =  \; NC_{\Phi} - AUT (A) \\
   = \; AUT (A) \cap {\mathcal{L}}_{COMP} (A)    \\
\end{multline}
\begin{multline}
  C_{\Phi} - DYN[(A , \omega)] \; =  \; NC_{\Phi} - DYN[(A , \omega)] \\
   = \; DYN[(A , \omega)] \cap {\mathcal{L}}_{COMP}(A)
\end{multline}
\textsl{\textbf{if and only if $ ( \, \, A \, , \, \omega \, ) $
is a noncommutative probability space of discrete type}}
\end{slide*}
\begin{slide*}
\section{Looking for Martin-L\"{o}f physically-quantum randomness: an issue of Algorithmic Free Probability Theory}
Given the unbiased noncommutative probability space $ (
L^{\infty} ( \Sigma_{NC}^{\infty} ) \; , \; \tau_{unbiased} ) $ of
the sequences on the one qubit noncommutative alphabet $
\Sigma_{NC} $ :
\begin{definition}
\end{definition}
UNARY PREDICATES ON  $ L^{\infty} ( \Sigma_{NC}^{\infty} ) $ :
\begin{multline}
  {\mathcal{P}} ( L^{\infty} ( \Sigma_{NC}^{\infty} )  ) \; \equiv \\
   \{ p_{\bar{x}} \, : \, \text{ predicate about } \\
    \bar{x} \in L^{\infty} ( \Sigma_{NC}^{\infty} )  \}
\end{multline}
\end{slide*}
\begin{slide*}
\begin{definition}
\end{definition}
$Q_{\Phi}$ - ALGORITHMICALLY TYPICAL PROPERTIES OF $ L^{\infty} (
\Sigma_{NC}^{\infty} ) $ :

$ Q_{\Phi}-{\mathcal{P}}  ( L^{\infty} ( \Sigma_{NC}^{\infty} )
)_{ALG-TYPICAL} \; \equiv $
\begin{multline}
  \{ \, p_{\bar{x}}  \in {\mathcal{P}} ( L^{\infty} ( \Sigma_{NC}^{\infty} )  )  \, :  \\
   \{ \bar{x} \in L^{\infty} ( \Sigma_{NC}^{\infty} ) \, : \, p_{\bar{x}} \text{ doesn't hold }
   \} \\
 \text{is a $Q_{\Phi}$-algorithmically null set} \}
\end{multline}
where \textbf{\emph{$Q_{\Phi}$ - ALGORITHMICALLY}} refers to
\textbf{computability by \emph{physical computers} obeying
\emph{Nonrelativistic or Partial Relativistic Quantum Mechanics}}
\end{slide*}
\begin{slide*}
\begin{definition} \label{def:random sequence of qubits}
\end{definition}
RANDOM SEQUENCES OF QUBITS :

$ Q_{\Phi}-RANDOM  ( L^{\infty} ( \Sigma_{NC}^{\infty} )  ) \;
\equiv $
\begin{multline}
   L^{\infty} ( \Sigma_{NC}^{\infty} )- \{ A \subset L^{\infty} ( \Sigma_{NC}^{\infty} )  \\
    \text{$Q_{\Phi}$- algorithmically null } \}
\end{multline}
\end{slide*}
\begin{slide*}
\begin{remark}
\end{remark}
WHAT LACKS TO COMPLETE DEFINITION\ref{def:random sequence of
qubits}

Clearly the definition\ref{def:random sequence of qubits} is
uncomplete until one gives the definition of $ Q_{\Phi}$-
algorithmically null subsets of $ L^{\infty} (
\Sigma_{NC}^{\infty} ) $.
\end{slide*}
\begin{slide*}
INGREDIENTS USEFUL TO IDENTIFY THE CORRECT NOTION OF  $
Q_{\Phi}$-ALGORITHICALLY NULL SUBSETS OF $ L^{\infty} (
\Sigma_{NC}^{\infty} ) $:
\begin{enumerate}
  \item the Pour - El Richards Theory
  \item the constraint\ref{con:free sequence of tosses of a quantum coin}
  \item the link exististing between \textbf{algorithmic comprimibility} and \textbf{probabilistic trasmission
  comprimibility} of a sequence of qubits
\end{enumerate}
\end{slide*}
\begin{slide*}
\begin{remark}
\end{remark}
WHAT POUR EL - RICHARDS THEORY CAN TELL ON THE COMPUTABILITY
THEORY OF THE SEQUENCES ON THE ONE QUBIT NONCOMMUTATIVE ALPHABET:

Since  $ ( L^{\infty} ( \Sigma_{NC}^{\infty} ) \; , \;
\tau_{unbiased} ) $ is not of discrete type Pour El Thesis can't
be advocated to identify $ {\mathcal{L}} ( L^{\infty} (
\Sigma_{NC}^{\infty} ) )_{COMP} $ and thus to construct
Algorithmic Information Theory on the sequences over $
\Sigma_{NC} $.

Anyway since an infinite chain of spin $ \frac{1}{2} $ at infinite
temperature is a \textbf{quantum physical system} described
exactly by the unbiased noncommutative probability space $ (
L^{\infty} ( \Sigma_{NC}^{\infty} ) \; , \; \tau_{unbiased} ) $ of
the sequences on the one qubit noncommutative alphabet $
\Sigma_{NC} $ it follows, looking at the example\ref{ex:one half
spin system}, that $ \otimes_{n \in  {\mathbb{N}}} \, \{ \,
{\mathbb{I}} \, , \, \sigma_{x} \, , \, \sigma_{y} \, ,
\,\sigma_{z} \, \} $ is an \textbf{effectively generating set} of
$ L^{\infty} ( \Sigma_{NC}^{\infty} ) $ and thus, for the
theorem\ref{th:of univocity}, individuates on it a
\textbf{computability structure}
\end{slide*}
\begin{slide*}
\begin{remark}
\end{remark}
NOT TRIVIALITY OF TRANSLATING CONSTRAINT\ref{con:free sequence of
tosses of a quantum coin} IN TERMS OF TYPICAL PROPERTIES

In the commutative case we saw that the
constraint\ref{con:independent sequence of tosses of a classical
coin} could simply be translated in terms of typical properties
as the constraint\ref{con:on classical algorithmic randomness}.

If the definition\ref{def:sequences on the noncommutative
alphabet} involved \textbf{free product}\cite{Hiai-Petz-00}
instead of \textbf{tensor products} of $W^{\star}$-algebras the
same would happen also for the constraint\ref{con:free sequence of
tosses of a quantum coin}, i.e. such a constraint could be simply
stated as:
\begin{constraint} \label{con:erroneous quantum algorithmic randomness}
\end{constraint}

\textsf{ERRONEOUS WAY OF LOOKING FOR THE DEFINITION OF $ RANDOM(
\Sigma_{NC}^{\infty} )$ }:

 \emph{the unary predicate $ p_{\bar{x}} \; \equiv
\; << \bar{x} \; \in  RANDOM( \Sigma_{NC}^{\infty} )   >> $ is a
$ Q_{\Phi}$-typical property of $\Sigma_{NC}^{\infty}$, i.e. $
p_{\bar{x}} \in Q_{\Phi}-{\mathcal{P}}  ( \Sigma_{NC}^{\infty}
)_{TYPICAL} $}
\end{slide*}
\begin{slide*}
Called $ c_{n} \in \Sigma $ the random variable on the
\textbf{unbiased probability space on the one cbit alphabet } $ (
\Sigma , C_{\frac{1}{2} , \frac{1}{2}} ) $ corresponding to the
result of the toss of a \textbf{classical coin} made at time $ n
\in { \mathbb{N}} $:
\begin{definition}
\end{definition}
NORMALIZED INDEPENDENT-LETTERS CLASSICAL INFORMATION SOURCE:

the $ \{ c _{n} \} $,  supposed to be an  \textbf{independent
sequence} on $ ( \Sigma , C_{\frac{1}{2} , \frac{1}{2}} ) $ so
that:

\begin{equation}
\begin{split}
  E ( c _{n} ) \; & = \; 0  \; \; \forall n \in {\mathbb{N}} \\
  E ( c _{n}^{2} ) \; & = \; 1  \; \; \forall n \in {\mathbb{N}}
\end{split}
\end{equation}
\end{slide*}
\begin{slide*}
An immediate argument of \textbf{Commutative Large Deviation
Theory} leads to \textbf{Shannon's Noiseless - Memoryless Coding
Theorem} \cite{Khinchin-57}, \cite{Billingsley-65},
\cite{Cover-Thomas-91}, \cite{Kakihara-99} implying that the
\textbf{probabilistic trasmission-comprimibility} for such a
classical information source is:
\begin{equation}
S_{Shannon}( C_{\frac{1}{2} , \frac{1}{2}}) \; = \; 1 \;
\frac{cbit}{letter}
\end{equation}
\end{slide*}
\begin{slide*}
Called $ c_{n} \in M_{2} ({\mathbb{C}}) $ the noncommutative
random variable on the \textbf{unbiased noncommutative
probability space on the one qubit alphabet} $( M_{2}
({\mathbb{C}})  \, , \, \tau_{2} ) $ corresponding to the result
of the toss of  a \textbf{quantum coin} made at time $ n \in {
\mathbb{N}} $:
\begin{definition}
\end{definition}
NORMALIZED INDEPENDENT-LETTERS QUANTUM INFORMATION SOURCE:

the $ \{ c _{n} \} $, supposed to be an  \textbf{independent
sequence} on $( M_{2} ({\mathbb{C}}) \, , \, \tau_{2} ) $ so that:
\begin{equation}
\begin{split}
  \tau_{2} ( c _{n} ) \; & = \; 0  \; \; \forall n \in {\mathbb{N}} \\
  \tau_{2} ( c _{n}^{2} ) \; & = \; 1  \; \; \forall n \in {\mathbb{N}}
\end{split}
\end{equation}
\end{slide*}
\begin{slide*}
\begin{definition}
\end{definition}
NORMALIZED FREE-LETTERS QUANTUM INFORMATION SOURCE:

the $ \{ c _{n} \} $, supposed to be a  \textbf{free sequence} on
$( M_{2} ({\mathbb{C}}) \, , \, \tau_{2} ) $ so that:
\begin{equation}
\begin{split}
  \tau_{2} ( c _{n} ) \; & = \; 0  \; \; \forall n \in {\mathbb{N}} \\
  \tau_{2} ( c _{n}^{2} ) \; & = \; 1  \; \; \forall n \in {\mathbb{N}}
\end{split}
\end{equation}
\end{slide*}
\begin{slide*}
\begin{remark}
\end{remark}
NOISELESS CODING THEOREM REGARDS THE INDEPENDENT-LETTERS QUANTUM
INFORMATION SOURCES AND NOT THE FREE-LETTERS QUANTUM INFORMATION
SOURCES

The \textbf{Noncommutative Large Deviation Theory}'s argument
\cite{Ohya-Petz-93}, \cite{Hiai-Petz-00} leading to
\textbf{Schumacher's Noiseless-Memoryless Quantum Coding Theorem}
\cite{Jozsa-97}, \cite{Schumacher-98}, \cite{Preskill-98},
\cite{Winter-99}, \cite{Petz-Mosonyi-99} implies that the
\textbf{probabilistic trasmission-comprimibility} of the
\textbf{normalized independent letters quantum information source}
is:
\begin{equation}
S_{Von \, Neumann}( \tau_{2} )  \; = \; 1 \; \frac{qubit}{letter}
\end{equation}
\end{slide*}
\begin{slide*}
\emph{But Schumacher's Theorem can't, obviously, be applied to the
\textbf{free-letters-quantum information source}} whose relevant
\textbf{large deviation theoretical entropy-functional} is
Voiculescu's \textbf{free entropy} \cite{Hiai-Petz-00}
\end{slide*}
\begin{slide*}
\begin{remark}
\end{remark}
COMMUTATIVE VERSUS NONCOMMUTATIVE LARGE DEVIATIONS FROM THE
CENTRAL LIMITS

The  conceptual meaning of the Noiseless Coding Theorem for any (
classical or quantum ) information source IS is:
\begin{itemize}
  \item the exponential decay of probability of large deviations from the \textbf{IS -
  central limit measure} $ P_{central} $ is governed by some \textbf{large deviation theoretical
  entropy-functional $ S_{IS} [ P ] $ }
\end{itemize}
\end{slide*}
\begin{slide*}
\begin{itemize}
  \item the conseguential possibility  of not-codifiying the $
  S_{IS}$ - not typical messages during the trasmission of information with asymptotically null
  misunderstanding-error
  \item the resulting   $ S_{IS}[P_{IS}]$ \textbf{probabilistic trasmission
  comprimibility} for IS
\end{itemize}
So it is important, first of all, to compare the Central Limit
Theorems of Commutative and Noncommutative Probability Theory
\end{slide*}
\begin{slide*}
\begin{theorem}\label{th:independent central limit}
\end{theorem}
CENTRAL LIMIT FOR THE NORMALIZED LETTERS-INDEPENDENT CLASSICAL
INFORMATION SOURCE

\begin{hypothesis}
$ \{ c _{n} \} $  letters-independent classical information source

$  m_{n} \; \equiv \; \frac{1}{\sqrt{n}} \sum_{k=1}^{n} c_{k} $

$ sup_{n} | E ( c_{n}^{k} ) | \; < \; + \infty \; \; \forall k
\in {\mathbb{N}} $
\end{hypothesis}
\begin{thesis}
$ meas-\lim_{n \rightarrow \infty} m_{n} \; = \; $
\textbf{standard gaussian measure}
\end{thesis}
\end{slide*}
\begin{slide*}
\begin{theorem}\label{th:free central limit}
\end{theorem}
CENTRAL LIMIT FOR THE NORMALIZED LETTERS-FREE QUANTUM INFORMATION
SOURCE

\begin{hypothesis}
$ \{ c _{n} \} $  letters-free quantum information source

$  m_{n} \; \equiv \; \frac{1}{\sqrt{n}} \sum_{k=1}^{n} c_{k} $

$ sup_{n} | \tau_{2} ( c_{n}^{k} ) | \; < \; + \infty  \; \;
\forall k \in {\mathbb{N}} $
\end{hypothesis}
\begin{thesis}
$ meas-\lim_{n \rightarrow \infty} m_{n} \; = \; $
\textbf{standard semicircle measure}
\end{thesis}
with:
\end{slide*}
\begin{slide*}
\begin{definition}
\end{definition}
GAUSSIAN  MEASURE OF MEAN m AND VARIANCE $\sigma^{2}$:

the probability measure on $ ( \, {\mathbb{R}} \, , \,
{\mathcal{F}}_{Borel} \, ) $ with density:
\begin{equation}
  g(m ,\sigma ; x ) \; \equiv \; \frac{1}{\sqrt{2 \pi \sigma^{2}}} e^{- \frac{ ( x - m )^{2}}{2 \sigma^{2}}}
\end{equation}
\begin{definition}
\end{definition}
STANDARD GAUSSIAN  MEASURE:

the probability measure on $ ( \, {\mathbb{R}} \, , \,
{\mathcal{F}}_{Borel} \, ) $ with density g(0 ,1; x )
\end{slide*}
\begin{slide*}
\begin{definition}
\end{definition}
SEMICIRCLE MEASURE OF MEAN m AND VARIANCE $ \frac{r^{2}}{4} $:

the probability measure on $ ( \, {\mathbb{R}} \, , \,
{\mathcal{F}}_{Borel} \, ) $ with density:

$ sc(m ,r ; x ) \; \equiv $
\begin{equation}
  \begin{cases}
    \frac{2}{\pi r^{2}}  \sqrt{ r^{2} - (x-m)^{2}} & \text{if $ m-r \leq x \leq m+r$}, \\
    0 & \text{otherwise}.
\end{cases}
\end{equation}
\begin{definition}
\end{definition}
STANDARD SEMICIRCLE MEASURE:

the probability measure on $ ( \, {\mathbb{R}} \, , \,
{\mathcal{F}}_{Borel} \, ) $ with density sc(0 ,2; x )
\end{slide*}
\begin{slide*}
MOMENTS OF THE STANDARD GAUSSIAN  MEASURE :

$ M_{n} \, [ g(0 ,1; x ) ] \; \equiv \int_{- \infty}^{+ \infty} dx
\, x^{n} \, g(0 ,1; x ) \; = $
\begin{equation}
  \begin{cases}
    ( 2k-1 ) \, !!  & \text{if $ n = 2k , k \in {\mathbb{N}}$}, \\
    0 & \text{otherwise}.
  \end{cases}
\end{equation}
MOMENTS OF THE STANDARD SEMICIRCLE MEASURE :

$ M_{n} \, [ sc(0 ,2; x ) ] \; \equiv \int_{- \infty}^{+ \infty}
dx \, x^{n} \, sc(0 ,2; x ) \; = $
\begin{equation}
  \begin{cases}
    \frac{1}{k+1} \, \begin{pmatrix}
      2 k  \\
      k \\
    \end{pmatrix}   & \text{if $ n = 2k , k \in {\mathbb{N}}$}, \\
    0 & \text{otherwise}.
  \end{cases}
\end{equation}
\end{slide*}
\begin{slide*}
\begin{remark}
\end{remark}
PROBABILISTIC ORIGIN  OF WIGNER'S THEOREM ON RANDOM MATRICES:

Random matrices belonging to the  Gaussian Unitary Ensemble are
asympotically-free random variables and conseguentially satisfy
the Free Central Limit Theorem resulting in Wigner's Theorem
\cite{Hiai-Petz-00},\cite{Mehta-91}
\end{slide*}
\begin{slide*}
Given a classical probability space $ ( \, \Omega \, , \, P \, )
$:
\begin{definition}
\end{definition}
NONCOMMUTATIVE PROBABILITY SPACE OF n $ \times $ n RANDOM MATRICES
W.R.T. $ ( \, \Omega \, , \, P \, ) $:

RANDOM-MATRICES$[ \, n \, , \, ( \, \Omega \, , \, P \, ) \, ] \;
\equiv \; ( A \, , \tau ) $ with:
\begin{multline}
   A \; \equiv  \; \{ X \,  n \times n \,  matrix  :  \\
  X_{ij} \in  L^{\infty} ( \, \Omega \, , \, P \, ) \\
   i,j = 1   , \ldots , n  \}
\end{multline}
$ \tau $ tracial state on A :
\begin{equation}
  \tau ( X ) \; \equiv \; \frac{1}{n} \, \sum_{i=1}^{n} \, E(X_{ii})
\end{equation}
\end{slide*}
\begin{slide*}
Given $ X \in RANDOM-MATRICES[ \, n \, , \, ( \, \Omega \, , \, P
\, ) \, ] $:
\begin{definition}
\end{definition}
EMPIRICAL EIGENVALUE DISTRIBUTION OF X :
\begin{equation}
  \mu_{emp} ( X ) \; \equiv \;  \frac{1}{n} \, \sum_{i=1}^{n}
  \delta( \lambda_{i} (X))
\end{equation}
\begin{definition}
\end{definition}
MEAN EIGENVALUE DISTRIBUTION OF X :
\begin{equation}
  \mu_{mean} ( X ) \; \equiv \; E (  \mu_{emp} ( X ) )
\end{equation}
where $ \lambda_{1}(X) \, , \ldots , \, \lambda_{n}(X) $ are the
(random) eigenvalues of X
\end{slide*}
\begin{slide*}
\begin{definition}
\end{definition}
n - DIMENSIONAL GAUSSIAN UNITARY ENSEMBLE :

$ GUE_{n} \; \equiv \; RANDOM-MATRICES[ \, n \, , \, ( \, \Omega
\, , \, P \, ) \, ] $ where $ ( \, \Omega \, , \, P \, ) $ is so
that given $ H \in GUE_{n} $ :
\begin{itemize}
  \item  $ H^{\dag} \; = \; H $ with probability one
  \item $ \{ \, \Re(H_{i j}) : i,j = 1 , \ldots , n \, \} \; \cup \; \{ \, \Im (H_{i j}) : i,j = 1 , \ldots , n \,
  \} $ is a family of independent Gaussian random variables
  \item
\begin{align}
  E ( H_{i j } ) & \; = \; 0  \; \; 1 \leq i \leq j \leq n  \\
  E ( H_{i j }^{2} )& \; = \; \frac{1}{n}  \; \; 1 \leq i \leq j \leq n \\
  E ( \Re ( H_{i j }^{2} ) ) & \; = \; E ( \Im ( H_{i j }^{2} ) ) \; = \;  \frac{1}{2 n}  \; \; 1 \leq i \leq j \leq n
\end{align}
\end{itemize}
\end{slide*}
\begin{slide*}

\end{slide*}
\end{document}